\documentclass[prd,aps,nofootinbib,showpacs,notitlepage,showkeys,preprintnumbers]
{revtex4-1}
\usepackage{graphicx,epsf,amsmath,amsfonts,amssymb,amsbsy}
\usepackage{epsfig}
\usepackage[mathscr]{eucal}
\newcommand{\ds}{\displaystyle}
\newcommand{\vev}[1]{\langle#1\rangle}
\newcommand{\mat}{\left ( \begin{array}}
\newcommand{\emat}{\end{array} \right )}
\newcommand{\vect}{\left ( \begin{array}{c}}
\newcommand{\evect}{\end{array} \right )}

\begin{document}

\title{Inhomogeneous charged pion condensation phenomenon in the
NJL$_2$ model with quark number and isospin chemical potentials}
\author{N.V. Gubina$^{1)}$, K.G. Klimenko$^{2)}$,
  S.G. Kurbanov$^{1)}$ and V.Ch. Zhukovsky$^{1)}$}
\affiliation{$^{1)}$ Faculty of
Physics, Moscow State University, 119991, Moscow, Russia}
\affiliation{$^{2)}$ IHEP and University "Dubna" (Protvino branch),
142281, Protvino, Moscow Region, Russia}

\begin{abstract}
The properties of two-flavored massive Nambu--Jona-Lasinio model in
(1+1)-dimensional spacetime are investigated in the presence of
isospin and quark number chemical potentials. The
consideration is performed in the large-$N_c$ limit, where $N_c$ is
the number of colored quarks. It is shown in the framework of this
model that charged pion condensation phenomenon of dense
quark/hadron isotopically asymmetric matter is rather a spatially
inhomogeneous than a homogeneous one.\end{abstract} \maketitle

\section{Introduction}

Recently,  much attention has been paid to the investigation of
the QCD phase diagram in terms of baryonic as well as isotopic
(isospin) chemical potentials. The reason is that dense baryonic
matter which can appear in heavy-ion collision experiments has an
evident isospin asymmetry. Moreover, the dense hadronic/quark matter
inside compact stars is also expected to be isotopically asymmetric.
To describe the above mentioned realistic situations, i.e. when the
baryonic density is comparatively low, usually different
nonperturbative methods or effective theories such as chiral
effective Lagrangians and especially  Nambu -- Jona-Lasinio (NJL)
type models \cite{njl} are employed. In this way, the QCD phase
diagram including chiral symmetry restoration
\cite{asakawa,ebert,sadooghi,hiller,boer}, color superconductivity
\cite{alford,klim,incera}, and charged pion condensation (PC)
phenomena \cite{son,ek,ak,mu,andersen} were investigated under
heavy-ion experimental and/or compact star conditions, i.e. in the
presence of such external conditions as temperature, chemical
potentials and possible external (chromo)magnetic fields (see the
above references).

Obviously, the (3+1)-dimensional NJL models depend on the cutoff
parameter which is typically chosen to be of the order of 1 GeV, so that
the results of their usage are valid only at {\it comparatively low
energies, temperatures and densities (chemical potentials)}.
Moreover, there exists also a class of renormalizable theories, the
(1+1)-dimensional chiral Gross--Neveu (GN) type models \cite{gn,ft},
\footnote{Below we shall use the notation ``NJL$_2$ model''  instead
of ``chiral GN model'' for (1+1)-dimensional models with {\it a
continuous chiral symmetry}, since the chiral structure of the
Lagrangian is the same as that of the (3+1)-dimensional NJL model.}
that can be used as a laboratory for the qualitative simulation of
specific properties of QCD at {\it arbitrary energies}.
Renormalizability, asymptotic freedom, as well as the spontaneous
breaking of chiral symmetry (in vacuum) are the most fundamental
inherent features both for QCD and all GN type models. In addition,
the $\mu-T$ phase diagram is qualitatively the same for the QCD and
GN models \cite{wolff,kgk1,barducci,chodos} (here $\mu$ is the quark
number chemical potential and $T$ is the temperature). Note also
that the GN type models are suitable for the description of physics
in quasi one-dimensional condensed matter systems like polyacetylene
\cite{caldas}. It is currently well understood (see, e.g., the
discussion in \cite{barducci,chodos,thies}) that the usual {\it no-go}
theorem \cite{coleman}, which generally forbids the spontaneous
breaking of any continuous symmetry in two-dimensional spacetime
does not work in the limit  $N_c\to\infty$, where $N_c$ is the
number of colored quarks. This follows from the fact that in the
limit of large $N_c$ the quantum fluctuations, which would otherwise
destroy a long-range order corresponding to a spontaneous symmetry
breaking, are suppressed by $1/N_c$ factors. Thus,  the effects
inherent for real dense quark matter, such as Cooper pairing phenomenon  (spontaneous breaking of the continuous $U(1)$
symmetry) or charged pion condensation (spontaneous breaking of the
continuous isospin symmetry) might be simulated in terms of a
simpler (1+1)-dimensional GN-type model, though only in the leading
order of the large $N_c$ approximation (see, e.g.,
\cite{chodos,abreu} and \cite{ektz,massive,ek2,gubina},
respectively).

This paper is devoted to investigation of the charged pion
condensation (PC) phenomenon in the framework of the (1+1)-dimensional
NJL model  with two quark flavors and in the presence of the quark
number ($\mu$) as well as isospin ($\mu_I$) chemical potentials.
The  consideration is performed in the leading order of the
$1/N_c$-expansion. In our previous papers \cite{ektz,massive,ek2}
the phase diagram of the above mentioned massless or massive NJL$_2$
model was already investigated in the case of homogeneous, i.e.
independent of space coordinate, order parameters (chiral and charged
pion condensates). The situation corresponds to the conserved
Lorentz and spatial translational invariance and is adequate to
physical systems in vacuum, i.e. at zero chemical potentials.
However, in dense baryonic matter, i.e. at nonzero quark number
chemical potential, there might appear new phases with a spatially
inhomogeneous chiral and/or charged pion condensates which destroy
both chiral and/or isospin as well as spatial translational
invariances of a system. In particular, the possibility of the phase
with inhomogeneous chiral condensate was discussed in the framework
of both (1+1)-dimensional \cite{thies,thies2,misha,gubina} and
(3+1)-dimensional
\cite{3+1,nakano,nickel,maedan,zfk,pisarski,basar,miransky} models.
At the same time the possibility of the phase with inhomogeneous
charged pion condensation is less investigated. \footnote{For
example, spatially inhomogeneous ansatz for the charged pion
condensate was investigated in the framework of the two-flavored
NJL$_4$ model in \cite{mu}. It was claimed there that inhomogeneous
PC phase is possible only at rather high values of an isotopic
chemical potential, $\mu_I>\Lambda$, where $\Lambda\sim 0.65$ MeV is
a cutoff parameter. So this result might be out of the scope of a
model application. Moreover, the authors of \cite{mu} made some
technical simplifications in their research of the inhomogeneous PC
phenomenon.}

Thus, in this paper,  in contrast to
\cite{ektz,massive,ek2}, we consider the phase portrait of the above
mentioned massive (1+1)-dimensional NJL model with two chemical
potentials, $\mu$ and $\mu_I$, in the leading order of the
$1/N_c$-expansion taking into account the possibility that the
charged pion condensate might become spatially inhomogeneous. The
temperature is taken to be zero. For simplicity, for the chiral
condensate we use a spatially homogeneous ansatz. (In contrast, in
our previous paper \cite{gubina}  the spatially inhomogeneous ansatz
for the chiral condensate and homogeneous one for the charged pion
condensate was used in the framework of the same massless NJL$_2$
model.) Notice once more, the isotopic asymmetry is an inevitable
property of dense quark matter which might be created in heavy-ion
collision experiments or inside compact stars. So, we believe that
such a simplified study in the framework of the two-dimensional NJL
model with isospin chemical potential could shed new light on the
properties of real dense baryonic matter and will provide a deeper
understanding of the charged PC phenomenon. In particular, it is
shown in our paper that in the framework of the model under
consideration a PC phase with {\it nonzero baryon density} is
realized just in the case of inhomogeneous charged pion condensate
but not in the case of homogeneous one. In analogy, one can expect
that in real (3+1)-dimensional dense hadronic/quark matter the
charged PC phenomenon is realized rather with inhomogeneous pion
condensate than with spatially homogeneous one.

The paper is organized as follows. In Section II we derive, in the
leading order of the large $N_c$-expansion, the general expression
for the thermodynamic potential of the two-flavored massive NJL$_2$
model with quark number chemical potential $\mu$ and isospin
chemical potential $\mu_I$ in the case of spatially homogeneous
chiral condensate and inhomogeneous PC. First, in Sec. III we
reduce our consideration to the case of homogeneous PC
and find that in this case only a PC phase with {\it zero density}
of quarks is possible in the model. Second, in Sec. IV it is shown
that charged PC phase with {\it nonzero quark density} in the
framework of the model is possible only with spatially inhomogeneous
charged pion condensate. Final Sec. V presents a summary and some
concluding remarks. The discussion of some technical problems are
relegated to two Appendices.

\section{ The model and its thermodynamic potential}
\label{effaction}

We consider a (1+1)-dimensional NJL$_2$ model to mimic the phase
structure of real dense quark matter composed of two massive quark
flavors ($u$- and $d$-  quarks). Its Lagrangian has the form:
\begin{eqnarray}
&&  L=\bar q\Big [\gamma^\rho\mathrm{i}\partial_\rho-m_0
+\mu\gamma^0+\frac{\mu_I}2 \tau_3\gamma^0\Big ]q+ \frac {G}{N_c}\Big
[(\bar qq)^2+(\bar q\mathrm{i}\gamma^5\vec\tau q)^2 \Big ],
\label{1}
\end{eqnarray}
where the quark field $q(x)\equiv q_{i\alpha}(x)$ is a flavor
doublet ($i=1,2$ or $i=u,d$) and color $N_c$-plet
($\alpha=1,...,N_c$) as well as a two-component Dirac spinor (the
summation in (\ref{1}) over flavor, color, and spinor indices is
implied); $\tau_k$ ($k=1,2,3$) are Pauli matrices; the quark number
chemical potential $\mu$ in (\ref{1}) is responsible for the nonzero
baryonic density of quark matter, whereas the isospin chemical
potential $\mu_I$ is taken into account in order to study properties
of quark matter at nonzero isospin densities (in this case the
densities of $u$ and $d$ quarks are different).
 The Dirac gamma matrices in two-dimensional spacetime
have the following form:
\begin{equation}
\begin{split}
\gamma^0=\sigma_1=
\begin{pmatrix}
0&1\\
1&0\\
\end{pmatrix};\qquad
\gamma^1=\mathrm{i}\sigma_2=
\begin{pmatrix}
0&1\\
-1&0\\
\end{pmatrix};\qquad
\gamma^5=\sigma_3=
\begin{pmatrix}
1&0\\
0&{-1}\\
\end{pmatrix}.
\end{split}
\end{equation}
Evidently, the
model (\ref{1}) is a simple generalization of the original
(1+1)-dimensional Gross-Neveu model \cite{gn} with a single massless
quark color $N_c$-plet to the case of two massive quark flavors and
additional chemical potentials. As a result, in the case under
consideration we have a modified flavor symmetry group, which depends
essentially on whether the bare quark mass $m_0$ and isospin chemical
potential $\mu_I$ take zero or nonzero values. Indeed, at $\mu_I
=0,m_0=0$ the Lagrangian (\ref{1}) is invariant under
transformations from the chiral $SU_L(2)\times SU_R(2)$ group.
Then, at $\mu_I \ne 0, m_0=0$ this symmetry is reduced to
$U_{I_3L}(1)\times U_{I_3R}(1)$, where $I_3=\tau_3/2$ is the third
component of the isospin operator (here and above the subscripts
$L,R$ mean that the corresponding group acts only on left, right
handed spinors, respectively). Evidently, this symmetry can also be
presented as $U_{I_3}(1)\times U_{AI_3}(1)$, where $U_{I_3}(1)$ is
the isospin subgroup and $U_{AI_3}(1)$ is the axial isospin
subgroup. Quarks are transformed under these subgroups as $q\to\exp
(\mathrm{i}\alpha\tau_3) q$ and $q\to\exp (\mathrm{i}
\alpha\gamma^5\tau_3) q$, respectively. In the case $m_0\ne
0,\mu_I=0$ the Lagrangian (\ref{1}) is invariant with respect to the
$SU_I(2)$, which is a diagonal subgroup of the chiral $SU_L(2)\times
SU_R(2)$ group. Finally, in the most general case with $m_0\ne
0,\mu_I\ne 0$ the initial model (\ref{1}) is symmetric under the
above mentioned isospin subgroup $U_{I_3}(1)$. In addition, in all
foregoing cases the model is invariant under color SU($N_c$)-, baryon
charge  $U_B(1)$- and electric charge $U_Q(1)$ groups. 

The linearized version of the Lagrangian (\ref{1}), which contains
composite bosonic fields $\sigma (x)$ and $\pi_a (x)$ $(a=1,2,3)$,
has the following form:
\begin{eqnarray}
\tilde L\ds &=&\bar q\Big [\gamma^\rho\mathrm{i}\partial_\rho-m_0
+\mu\gamma^0+ \frac{\mu_I}2\tau_3\gamma^0-\sigma
-\mathrm{i}\gamma^5\pi_a\tau_a\Big ]q
 -\frac{N_c}{4G}\Big [\sigma\sigma+\pi_a\pi_a\Big ].
\label{2}
\end{eqnarray}
From the Lagrangian (\ref{2}) one obtains the following constraint
equations for the bosonic fields
\begin{eqnarray}
\sigma(x)=-2\frac G{N_c}(\bar qq);~~~\pi_a (x)=-2\frac G{N_c}(\bar q
\mathrm{i}\gamma^5\tau_a q). \label{200}
\end{eqnarray}
Obviously, the Lagrangian (\ref{2}) is equivalent to the Lagrangian
(\ref{1}) when using the constraint equations (\ref{200}).
Furthermore, it is clear that the bosonic fields (\ref{200}) are
transformed under the isospin $U_{I_3}(1)$ subgroup in the
following manner:
\begin{eqnarray}
U_{I_3}(1):~~~&&\sigma\to\sigma;~~\pi_3\to\pi_3;~~\pi_1\to\cos
(2\alpha)\pi_1+\sin (2\alpha)\pi_2;~~\pi_2\to\cos
(2\alpha)\pi_2-\sin (2\alpha)\pi_1, \label{201}
\end{eqnarray}
i.e the expression ($\pi_1^2+\pi_2^2$) remains unchanged under
transformations  of the isospin subgroup $U_{I_3}(1)$.

To avoid the {\it no-go} theorem \cite{coleman}, which forbids the
spontaneous breaking of 
continuous symmetries in the considered
case of one space dimension, we restrict the discussion only to the
leading order of the large $N_c$ expansion
(i.e. to the case $N_c\rightarrow\infty$), where this theorem is not
valid \cite{barducci,chodos,thies}. In particular, the effective
action $S_{\mathrm{eff}}(\sigma,\pi_a)$ can be found in this
approximation through the relation:
\begin{eqnarray}
\exp(\mathrm{i} {\cal S}_{\rm {eff}}(\sigma,\pi_a))=
  N'\int[d\bar q][dq]\exp\Bigl(\mathrm{i}\int\tilde  L\,d^2
  x\Bigr),\label{A1}
\end{eqnarray}
where $N'$ is a normalization constant. It is clear from (\ref{2})
and (\ref{A1}) that
\begin{eqnarray}
&&{\cal S}_{\rm {eff}} (\sigma,\pi_a)
=-N_c\int\frac{\sigma^2+\pi^2_a}{4G}d^2x+ \tilde {\cal S}_{\rm
{eff}}, \label{A2}
\end{eqnarray}
where the quark contribution to the effective action, i.e. the term
$\tilde {\cal S}_{\rm {eff}}$ in (\ref{A2}), is given as follows
\begin{equation}
\exp(\mathrm{i}\tilde {\cal S}_{\rm {eff}})=N'\int [d\bar
q][dq]\exp\Bigl(\mathrm{i}\int\bar q [\mathrm{i}\gamma^\rho\partial_\rho- m_0+\mu\gamma^0+
\nu\tau_3\gamma^0-\sigma -\mathrm{i}\gamma^5\pi_a\tau_a] q~d^2 x\Bigr).
\label{A3}
\end{equation}
Here we used the notation $\nu=\mu_I/2$. The ground state expectation
values  $\vev{\sigma(x)}$ and
$\vev{\pi_a(x)}$ of the composite bosonic fields are determined by
the saddle point equations,
\begin{eqnarray}
\frac{\delta {\cal S}_{\rm {eff}}}{\delta\sigma (x)}=0,~~~~~
\frac{\delta {\cal S}_{\rm {eff}}}{\delta\pi_a (x)}=0,~~~~~
\label{5}
\end{eqnarray}
where $a=1,2,3$. In vacuum, i.e. in the state corresponding to an
empty space with zero particle density and zero values of the
chemical potentials $\mu$ and $\mu_I$, the quantities
$\vev{\sigma(x)}$ and $\vev{\pi_a(x)}$ do not depend on space
coordinates. However, in dense quark medium, when $\mu\ne 0$, $\mu_I\ne
0$, the ground state expectation values of bosonic fields might have a
nontrivial dependence on $x$. In particular, in this paper we will
use the following ansatz:
\begin{eqnarray}
\vev{\sigma(x)}=M-m_0,~~~\vev{\pi_3(x)}=0,~~~\vev{\pi_1(x)}=\Delta\cos(2bx),~~~ \vev{\pi_2(x)}=\Delta\sin(2bx), \label{6}
\end{eqnarray}
where $M,b$, and $\Delta$ are constant quantities. In fact, they are
coordinates of the global minimum point of the thermodynamic potential (TDP) $\Omega (M,b,\Delta)$.
\footnote{Here and in what follows we will use a rather conventional
  notation
"global" minimum in the sense that among all our numerically found
local minima the thermodynamical potential takes in their case the
lowest value. This does not exclude the possibility that there exist
other inhomogeneous condensates, different from (\ref{6}), which
lead to ground states with even lower values of the TDP.}
In the leading order of the large $N_c$-expansion it is defined by the
following expression:
\begin{equation}
\int d^2x \Omega (M,b,\Delta)=-\frac{1}{N_c}{\cal S}_{\rm {eff}}\big (\sigma(x),\pi_a(x)\big )\Big|_{\sigma (x)=\vev{\sigma(x)},\pi_a(x)=\vev{\pi_a(x)}} ,
\end{equation}
which gives
\begin{equation}
\int d^2x\Omega (M,b,\Delta)\,\,=\,\,\int
d^2x\frac{(M-m_0)^2+\Delta^2}{4G}+\frac{\mathrm{i}}{N_c}\ln\left (
\int [d\bar q][dq]\exp\Bigl(\mathrm{i}\int d^2
x\bar q {\cal D} q \Bigr)\right ),
\label{8}
\end{equation}
where
\begin{equation}
\bar q {\cal D} q=\bar q\big (\gamma^\rho\mathrm{i}\partial_\rho +\mu\gamma^0+
\nu\tau_3\gamma^0-M\big )q-\Delta\big (\bar q_u\mathrm{i}\gamma^5 q_d\big )e^{-2\mathrm{i}bx}-\Delta\big (\bar q_d\mathrm{i}\gamma^5 q_u\big )e^{2\mathrm{i}bx}.\label{9}
\end{equation}
(Remember, in this formula $q$ is indeed a flavor doublet, i.e.
$q=(q_u,q_d)^T$.) To proceed, let us introduce in
(\ref{8})-(\ref{9}) the new quark doublets, $\psi$ and $\bar\psi$,
namely: $\psi=\exp(\mathrm{i}\tau_3bx)q$ and $\bar\psi = \bar
q\exp(-\mathrm{i}\tau_3bx)$. Since this transformation of quark
fields does not change the path integral measure in (\ref{8}), the
expression (\ref{8}) for the thermodynamic potential is easily
transformed to the following one:
\begin{eqnarray}
\int d^2x\Omega (M,b,\Delta)&=&\int
d^2x\frac{(M-m_0)^2+\Delta^2}{4G}+\frac{\mathrm{i}}{N_c}\ln\left (
\int [d\bar\psi][d\psi]\exp\Bigl(\mathrm{i}\int d^2
x\bar\psi D\psi \Bigr)\right ),
\label{10}
\end{eqnarray}
where  instead of the $x-$dependent Dirac operator ${\cal D}$ a new
$x-$independent operator $D$ appears
\begin{equation}
D=\gamma^\nu \mathrm{i}\partial_\nu -M +\mu\gamma^0+\tau_3\gamma^1b+\nu\tau_3\gamma^0-\mathrm{i}\Delta\tau_1\gamma^5.
\label{110}
\end{equation}
The expression (\ref{10}) for the thermodynamic potential is easily
transformed to the following one:
\begin{eqnarray}
\Omega (M,b,\Delta)&=&\frac{(M-m_0)^2+\Delta^2}{4G}+\mathrm{i}\frac{{\rm
Tr}_{sfx}\ln D}{N_c\int d^2x}\nonumber\\
&=&\frac{(M-m_0)^2+\Delta^2}{4G}+\mathrm{i}{\rm
Tr}_{sf}\int\frac{d^2p}{(2\pi)^2}\ln\overline{D}(p),
\label{11}
\end{eqnarray}
where the Tr-operation ${\rm Tr}_{sfx}$ stands for the trace in
spinor- ($s$), flavor- ($f$) as well as two-dimensional coordinate-
($x$) spaces, respectively, and ${\rm Tr}_{sf}$ is the respective
trace without $x-$space. Moreover, $\overline{D}(p)=\not\!p
+\mu\gamma^0+ \tau_3\gamma^1 b+\nu\tau_3
\gamma^0-M-\mathrm{i}\gamma^5\Delta\tau_1$ is the momentum space
representation of the Dirac operator $D$ (\ref{110}). Obviously,
$\overline{D}(p)$ is a 4$\times$4 matrix in the direct product of
the spinor and flavor spaces. Since
$\mathrm{Tr}_{sf}\mathrm{ln}\overline{D}(p)=\ln\det
\overline{D}(p)$, one can  evaluate the expression  (\ref{11}) with
a help of any program of analytical calculations and find
\begin{eqnarray}
\Omega (M,b,\Delta)\equiv\Omega^{un} (M,b,\Delta)
&=&\frac{(M-m_0)^2+\Delta^2}{4G}+\mathrm{i}\int\frac{d^2p}{(2\pi)^2}
\ln\det \overline{D}(p), \label{12}
\end{eqnarray}
where
\begin{eqnarray}
\det\overline{D}(p)&=&\Delta^4+2\Delta^2(M^2+p_1^2+\nu^2-b^2-\eta^2)\nonumber\\
&+&\big (M^2+(p_1-b)^2-(\eta+\nu)^2\big )\big
(M^2+(p_1+b)^2-(\eta-\nu)^2\big )  \label{26}
\end{eqnarray}
and $\eta=p_0+\mu$. (In order to emphasize the  fact that the
expression (\ref{12}) is divergent, i.e. unrenormalized,  we
use in this TDP notation the superscript ``un``.) Obviously, the
function $\Omega^{un} (M,b,\Delta)$ is symmetric with respect to the
transformation
$\Delta\to -\Delta$. (At $m_0=0$ it is also invariant  with respect
to the  $M\to-M$ transformation.) Moreover, it is invariant under
each of the transformations $b\to-b$, $\mu\to-\mu$ and $\nu\to-\nu$.
\footnote{Indeed, if simultaneously with $b\to-b$ or $\mu\to-\mu$
transformations we perform in the integral (\ref{12}) the  $p_1\to
-p_1$ or $p_0\to -p_0$ change of variables, respectively, then one
can easily see that the expression (\ref{12}) remains intact.
Finally, if $\nu\to-\nu$, we should transform $p_1\to -p_1$ in the
integral (\ref{12}) in order to be convinced that the TDP remains
unchanged. } Hence, without loss of generality, we restrict
ourselves to the constraints  $\Delta\ge 0$, $\mu\ge 0$, $b\ge 0$,
and $\nu\ge 0$.  In the following, we will investigate the global
minimum point of the TDP (\ref{12}) just on this region. However, first
of all let us consider the case of spatially homogeneous
condensates, i.e. the $b=0$ case.

\section{The case of homogeneous charged pion condensate, $b=0$}
\label{hom}

Supposing that $b=0$ in (\ref{12}), we obtain after some technical
calculations the TDP for the case of spatially homogeneous charged
pion condensate,
\begin{eqnarray}
\Omega^{un}(M,\Delta)
=\frac{(M-m_0)^2+\Delta^2}{4G}+\mathrm{i}\int\frac{d^2p}{(2\pi)^2}\ln
\Big\{\Big [(p_0+\mu)^2-(E^+_{\Delta})^2\Big ]\Big
[(p_0+\mu)^2-(E^-_{\Delta})^2\Big ]\Big\}, \label{120}
\end{eqnarray}
where
\begin{eqnarray}
E_\Delta^\pm=\sqrt{(E^\pm)^2+\Delta^2},~~~ E^\pm=E\pm \nu,~~~\nu=\mu_I/2,~~~
E=\sqrt{p_1^2+M^2}. \label{13}
\end{eqnarray}
The argument of the $\ln (x)$-function in (\ref{120}) is proportional
to the inverse quark propagator in the energy-momentum space
representation. Hence,
its zeros are the poles of the quark propagator. So, using (\ref{120})
one can find
the dispersion laws for quasiparticles, i.e. the momentum dependence
of the quark ($p_{0u}$, $p_{0d}$) and antiquark ($p_{0\bar u}$,
$p_{0\bar d}$) energies, in a medium (the full expression of the
quark propagator matrix is presented in Appendix B of paper
\cite{massive}):
\begin{equation}
p_{0u}=E_\Delta^--\mu,~~~p_{0d}=E_\Delta^+-\mu,~~
p_{0\bar u}=-(E_\Delta^++\mu),~~ p_{0\bar d}=-(E_\Delta^-+\mu).
\label{E6}
\end{equation}
Integrating in (\ref{120}) over $p_0$ (see in
\cite{ektz} for similar integrals), one obtains for the unrenormalized
TDP of the system at zero temperature the following expression:
\begin{eqnarray}
\Omega^{un}
(M,\Delta)&=&\frac{(M-m_0)^2+\Delta^2}{4G}-\int_{-\infty}^{\infty}
\frac{dp_1}{2\pi}
\Big\{E_\Delta^++E_\Delta^-\nonumber\\
&+&(\mu-E_\Delta^+)\theta(\mu-E_\Delta^+)+(\mu-E_\Delta^-)\theta(\mu-E_\Delta^-)\Big\},
\label{14}
\end{eqnarray}
where $\theta(x)$ is the Heaviside theta-function. It is clear that
the TDP (\ref{14}) is an ultraviolet divergent quantity, so in order
to get any physical information one should renormalize it, using a
special dependence of such quantities as the bare coupling
constant $G$ and the bare quark mass $m_0$ on the cutoff parameter
$\Lambda$ ($\Lambda$ restricts the integration region in the
divergent integral in (\ref{14}), $|p_1|<\Lambda$). The
renormalization procedure for the simplest massive GN model was
already discussed in the literature, see, e.g., in
\cite{kgk1,barducci,massive}. In a similar way, it is easy to see
that, cutting of the divergent integral in (\ref{14}) and then using
the substitution $G\equiv G(\Lambda)$ and $m_0\equiv mG(\Lambda)$,
where
\begin{eqnarray}
\frac{1}{2G(\Lambda)}=\frac{2}{\pi}\ln\left
(\frac{2\Lambda}{M_0}\right ) \label{16}
\end{eqnarray}
and $m,M_0$ are new free finite renormalization group invariant
massive parameters\footnote{\label{foot1} Note, the quantity $m$ is
not equal to the physical, or dynamical, quark mass $M$. The last
one is defined by the pole position of the quark propagator.
Alternatively, it can be found as a gap, i.e. one of the coordinates
of the global minimum point of the thermodynamic potential. However,
parameter $M_0$ is equal to a dynamically generated quark mass in
the vacuum and at $m_0=0$ (a more detailed discussion on the
physical essence of these parameters is given in \cite{massive}).}
(which do not depend on the cutoff $\Lambda$), it is possible to
obtain in the limit $\Lambda\to\infty$ a finite renormalization
group invariant expression for the TDP (for details see, e.g., the
papers \cite{massive}). Namely,
\begin{eqnarray}
\Omega^{ren}(M,\Delta)&&
=V_0(M,\Delta)-\frac{mM}{2}-\int_{-\infty}^{\infty}
\frac{dp_1}{2\pi}\Big\{E^+_{\Delta}+E^-_{\Delta}-2\sqrt{p_1^2+M^2+
\Delta^2}\nonumber\\
&&+(\mu-E^+_{\Delta})
\theta(\mu-E^+_{\Delta})+(\mu-E^-_{\Delta})\theta
(\mu-E^-_{\Delta})\Big\}, \label{22}
\end{eqnarray}
where
\begin{eqnarray}
V_0(M,\Delta)
=\frac{M^2+\Delta^2}{2\pi}\left [\ln\left
(\frac{M^2+\Delta^2}{M_0^2}\right )-1\right ] \label{23}
\end{eqnarray}
is a finite renormalization group invariant expression for the TDP
(\ref{22}) in vacuum, i.e. at $\mu=0$ and $\mu_I=0$,  taken in the
chiral limit, i.e. at $m=0$.

In the following, when studying the phase structure, the quantity
$M_0$ is still treated as a free parameter, however instead of the
massive parameter $m$ of the model we will use a dimensionless
parameter, $\tilde\alpha\equiv\pi m/M_0$. As a result, one can see
that in the massive NJL$_2$ model the dimensional transmutation
effect is absent formally. Indeed, both before and after
renormalization this massive model is parameterized by one massive-
and one dimensionless quantity. (Before renormalization the model is
characterized by a bare mass $m_0$ and a dimensionless bare coupling
constant $G$, while after renormalization the mass $M_0$ and
dimensionless quantity $\tilde\alpha$ are free model parameters.) In
contrast, in the massless GN-type models, i.e. at $m_0=0$, the
coupling constant $G$ is replaced after renormalization by the
massive parameter $M_0$ (it is the so-called dimensional
transmutation phenomenon).

In our subsequent calculations throughout the paper the quantity
$\tilde\alpha$ is fixed by $\tilde\alpha=\tilde\alpha_0\approx
0.17$. In this case we have in the initial NJL$_2$ model the same
relation between the pion mass and the dynamical quark mass in
vacuum as in some NJL-type models in the realistic case of the
(3+1)-spacetime \cite{massive}.

Investigating the behavior of the global minimum point (whose
coordinates are  just the gaps $M$ and $\Delta$) of the TDP
(\ref{22}) vs chemical potentials, it is possible to establish the
phase structure presented in Fig. 1. There, in the phases 1, 2 and 3
the gap $\Delta$ is vanishing, i.e. these are the normal quark matter
phases
with a nonzero gap $M$. However, at the boundaries between phases the
gap $M$ changes its value by a jump (the details of the
investigation, including the behavior of gaps, particle densities,
meson masses etc, are presented in \cite{massive}). For each point
$(\nu,\mu)$ of the vacuum region of Fig. 1 we have $\Delta=0$ and
$M\approx 1.04M_0$
(the physical meaning of the parameter $M_0$ is
described in footnote \ref{foot1}).
Finally, one can see in Fig. 1 the
homogeneous charged pion condensation phase in which both gaps are
not equal to zero. What is more interesting for us is that all over
this phase the quark number density  $n_q=-\partial\Omega^{ren}
(M,\Delta)/\partial\mu$ is equal to zero, $n_q=0$.

Therefore, in dense (i.e. with nonzero $n_q$) quark matter,
mimicked by the initial GN-type model, the phase with spatially
homogeneous charged pion condensation can not be realized.
\begin{figure}
\includegraphics[width=0.45\textwidth]{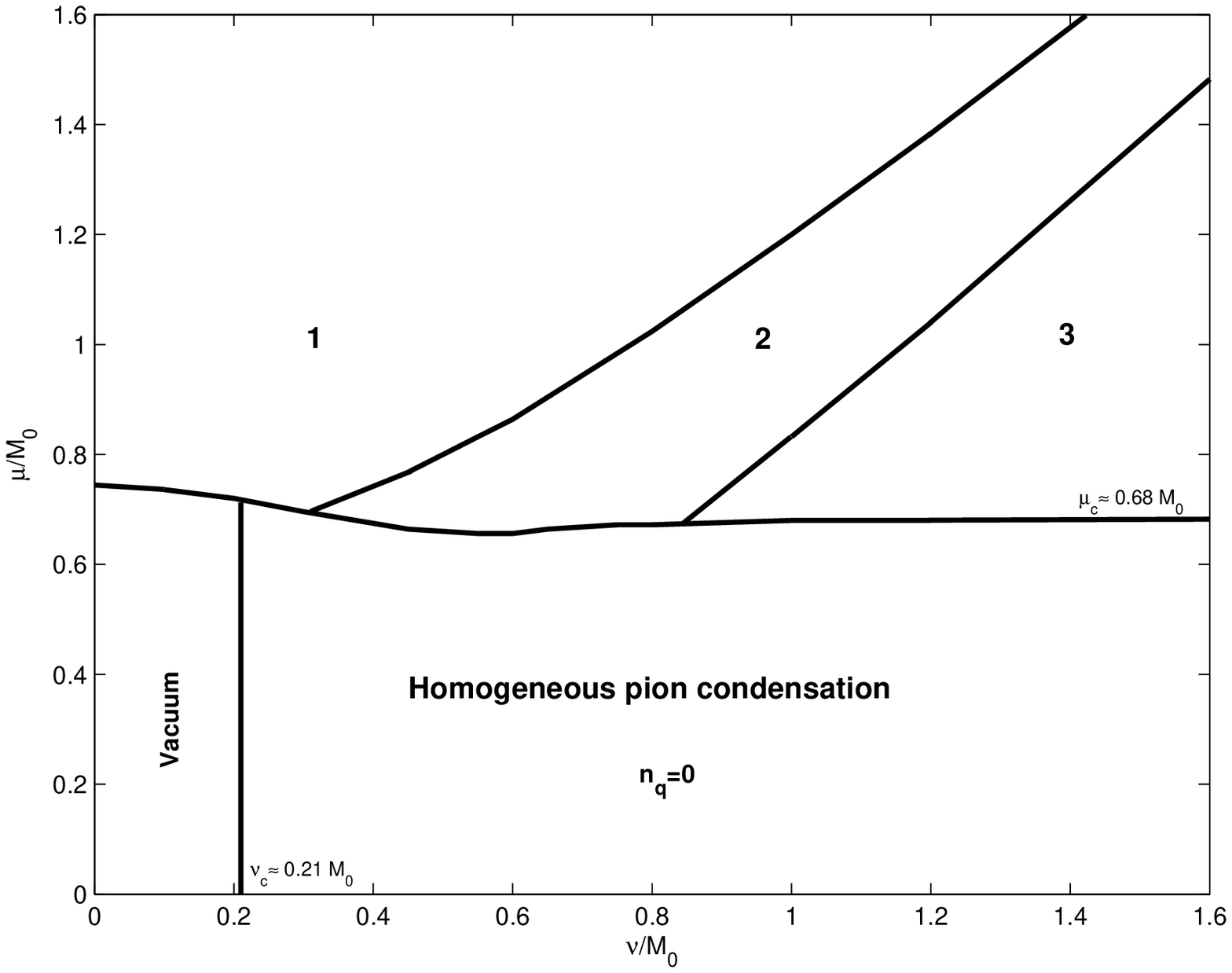}
\hfill
\includegraphics[width=0.45\textwidth]{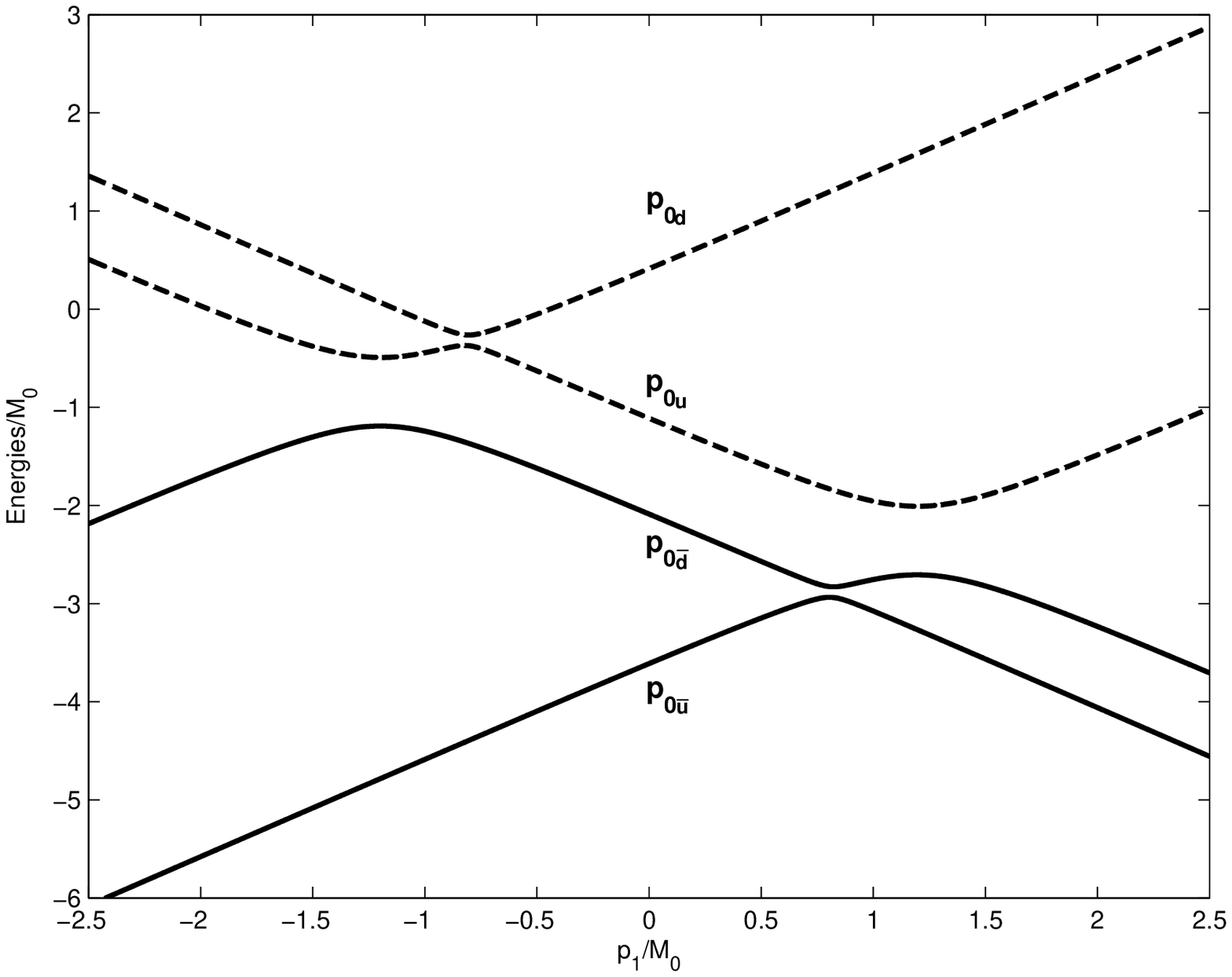}
\\
\parbox[t]{0.45\textwidth}{
\caption{The $(\mu,\nu)$ phase portrait of the model
considered at $T=0$ and $\nu\equiv\mu_I/2>0$ in the case of spatially
homogeneous condensates. Numbers 1, 2 and 3 denote different
normal quark matter phases with $\Delta=0$, $M\ne 0$. On all the lines
of the Figure, first order phase transitions occur except for the
boundary
between vacuum and homogeneous pion condensation phase, where a second
order phase transition takes place. $n_q$ is the quark number
density. } } \hfill
\parbox[t]{0.45\textwidth}{
\caption{The quasiparticle energies (\ref{E61}) vs $p_1$ at
$\mu=0.8M_0$, $\nu=1.2M_0$, $\Delta=0.35M_0$, $M=0.06M_0$, and $b=0.76M_0$. } }
\end{figure}

\section{Inhomogeneous ansatz for the charged pion condensate, $b\ne 0$}
\label{inhom}

In this Section, the possibility for the ground state of
the initial NJL$_2$ model (1) with spatially inhomogeneous charged
pion condensate is investigated. We start with the unrenormalized
TDP (\ref{12}) which can be rewritten in the form
\begin{eqnarray}
\Omega^{un} (M,b,\Delta)
&=&\frac{(M-m_0)^2+\Delta^2}{4G}+\mathrm{i}\int\frac{d^2p}{(2\pi)^2}
\ln\big ( \eta^4 + A\eta^2 + B\eta + C\big ), \label{012}
\end{eqnarray}
where the notation $\eta = p_0+\mu$ as well as the identity
$\det\overline{D}(p)\equiv\eta^4 + A\eta^2 + B\eta + C$ with
\begin{eqnarray}
\nonumber
A&=&-2(M^2+b^2+p_1^2+\nu^2+\Delta^2),~~~~B=-8p_1 b \nu,\nonumber\\
C&=&(M^2+b^2+p_1^2+\nu^2+\Delta^2)^2-4(p_1^2\nu^2
+ b^2\nu^2 + \Delta^2b^2 + M^2\nu^2 +p_1^2b^2)\label{1200}
\end{eqnarray}
are used. The argument of the $\ln$-function in (\ref{012}) can be
expanded into a product of four linear multipliers,
\begin{eqnarray}
\Omega^{un} (M,b,\Delta)
&=&\frac{(M-m_0)^2+\Delta^2}{4G}+\mathrm{i}\int\frac{d^2p}{(2\pi)^2}\ln\left
[ 
(\eta-\eta^{+}_1)(\eta-\eta^{-}_1 )(\eta-\eta^{+}_2)(\eta-\eta^{-}_2 )\right ],
\label{013}
\end{eqnarray}
where $\eta^{\pm}_k$ are presented in Appendix \ref{ApA} by the
expressions (\ref{AA1}). If the quantities $A,B$ and $C$ in
(\ref{012}) are defined by the relations (\ref{1200}), then
numerical analysis shows that all the roots $\eta^{\pm}_k$ are
real valued quantities vs gaps $M,\Delta,b$, chemical potentials
$\mu,\mu_I$ and spatial momentum $p_1$. Moreover, two of the roots, $\eta^{\pm}_2$, are negative valued quantities. Taking into account the
remark after formula (\ref{13}), it is possible to obtain
immediately from (\ref{013}) the quark-antiquark dispersion laws,
\begin{equation}
p_{0u}=\eta_1^--\mu,~~~p_{0d}=\eta_1^+-\mu,~~
p_{0\bar u}=\eta_2^--\mu,~~ p_{0\bar d}=\eta_2^+-\mu.
\label{E61}
\end{equation}
Note that at $b=0$ the quasiparticle energies (\ref{E61}) coincide
with the corresponding expressions from (\ref{E6}). At the
particular values of the chemical potentials and gaps the plots of
the quasiparticle energies $p_{0u}$,... (\ref{E61}) vs $p_1$ are
presented in Fig. 2.

Now, it is possible to perform the $p_0$-integration in (\ref{013})
using the general formula (see Appendix \ref{ApB})
\begin{eqnarray}
\int_{-\infty}^\infty dp_0\ln\big (p_0-a)=\mathrm{i}\pi|a|,\label{int}
\end{eqnarray}
where $a$ is a real quantity. As a result, we have
\begin{eqnarray}
\Omega^{un} (M,b,\Delta)
&=&\frac{(M-m_0)^2+\Delta^2}{4G}-\int_{-\infty}^\infty\frac{dp_1}{4\pi}\left [
|\mu-\eta^{+}_1|+|\mu-\eta^{-}_1|+|\mu-\eta^{+}_2|+|\mu-\eta^{-}_2|\right ].
\label{15}
\end{eqnarray}
To renormalize the expression (\ref{15}) we must first regularize
it. In this connection, it is necessary to make the following
remark. In the case of homogeneous condensates (see the previous
Section) usually the momentum cutoff regularization scheme is
used. However it does not work in the case of spatially inhomogeneous
condensates since three-momentum is no longer conserved\footnote{If
the momentum cutoff regularization is used in the inhomogeneous case, the TDP acquires some non-physical properties such as unboundedness from below 
with respect to $b$, etc. As a result, an additional modification of the TDP is needed (for details see in \cite{mu,gubina,miransky}).}. As discussed in the recent papers \cite{gubina,nakano,nickel,zfk}, an adequate
regularization scheme in the case of spatially inhomogeneous phases
is that with an energy constraint equal for all quasiparticles. So, dealing with spatial inhomogeneity, one can use, e.g., the Schwinger proper-time regularization, dimensional regularization etc. In particular, in our
recent paper \cite{zfk} the {\it symmetric energy cutoff regularization} scheme was proposed in considering the behavior of chiral density waves in the presence of an external magnetic field in the framework of 
the four-dimensional Nambu--Jona-Lasinio model. There, for each quasiparticle the same (finite) interval of their energy values was allowed to contribute to the regularized thermodynamic potential. In the present
investigation we will also use the symmetric energy cutoff regularization 
for the TDP, i.e.  
 \begin{eqnarray}
\Omega^{un} (M,b,\Delta)&\longrightarrow&\Omega^{reg} (M,b,\Delta)=
\frac{(M-m_0)^2+\Delta^2}{4G}\nonumber\\
&-&\int_{-\infty}^\infty\frac{dp_1}{4\pi}\big [
|p_{0u}|\theta(\Lambda-|p_{0u}|)+|p_{0d}|\theta(\Lambda-|p_{0d}|)+|p_{0\bar u}|\theta(\Lambda-|p_{0\bar u}|)+|p_{0\bar d}|\theta(\Lambda-|p_{0\bar d}|)\big ],
\label{20}
\end{eqnarray}
where the notations (\ref{E61}) for quasiparticle energies $p_{0u}$
etc. are used. Now, let us consider the identity  
\begin{eqnarray}
\Omega^{reg} (M,b,\Delta)=\left (\Omega^{reg} (M,b,\Delta)-\Omega^{reg}
(M,b,\Delta)\big |_{b=0,\mu=0,\nu=0}\right )+\Omega^{reg}
(M,b,\Delta)\big |_{b=0,\mu=0,\nu=0}. \label{17}
\end{eqnarray}
Clearly, at $b=0,\mu=0,\nu=0$ for all  quantities $\eta^{\pm}_{1,2}$
one finds the  relation $|\eta^{\pm}_{1,2}|=\sqrt{p_1^2+M^2+\Delta^2}$,
so that
\begin{eqnarray}
\Omega^{reg} (M,b,\Delta)\big |_{b=0,\mu=0,\nu=0}&=&\frac{(M-m_0)^2+\Delta^2}{4G}-\int_{-\infty}^\infty\frac{dp_1}{\pi}\sqrt{p_1^2+M^2+\Delta^2}~\theta\left (\Lambda-\sqrt{p_1^2+M^2+\Delta^2}\right ).
\label{18}
\end{eqnarray}
Since the expression in parenthesis in (\ref{17}) is an
ultraviolet (UV) convergent one, i.e. it is a finite quantity in the
$\Lambda\to \infty$ limit, we see that  in (\ref{17}) all the UV
divergences are located in the last term which is nothing but energy
cutoff regularized vacuum thermodynamic potential of the system
(\ref{18}). Hence, in order to renormalize the TDP $\Omega^{un}
(M,b,\Delta)$ (\ref{15}) it is sufficient to remove UV divergences
from the quantity (\ref{18}) by substituting  in 
(\ref{18}) $G\equiv G(\Lambda)$ and $m_0=mG(\Lambda)$ 
by quantities with an appropriate behavior of $G(\Lambda)$ vs $\Lambda$. As a result, we have
\begin{eqnarray}
&&\Omega^{un} (M,b,\Delta)\longrightarrow\Omega^{ren} (M,b,\Delta)=
V_0(M,\Delta)-\frac{mM}{2}
-\lim_{\Lambda\to\infty}\bigg\{\int_{-\infty}^\infty\frac{dp_1}{4\pi}\bigg [|p_{0u}|\theta(\Lambda-|p_{0u}|)+|p_{0d}|\theta(\Lambda-|p_{0d}|)
\nonumber\\
&&~~~+|p_{0\bar u}|\theta(\Lambda-|p_{0\bar u}|)+|p_{0\bar d}|\theta(\Lambda-|p_{0\bar d}|)-4\sqrt{p_1^2+M^2+\Delta^2}~\theta\left (\Lambda-\sqrt{p_1^2+M^2+\Delta^2}\right )\bigg ]\bigg\}. \label{21}
\end{eqnarray}
where $V_0(M,\Delta)$ is given in (\ref{23}).
\begin{figure}
\includegraphics[width=0.45\textwidth]{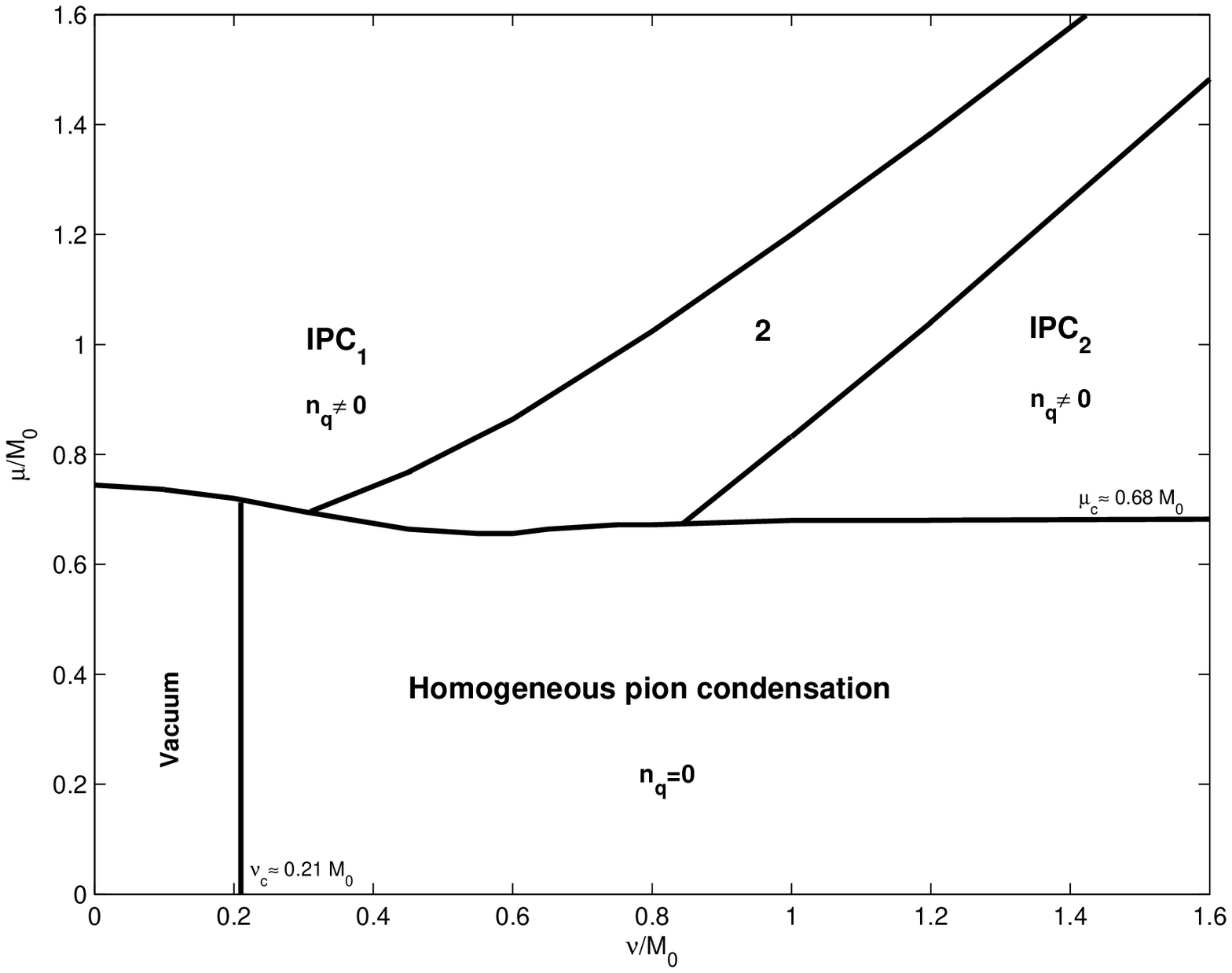}
\hfill
\includegraphics[width=0.45\textwidth]{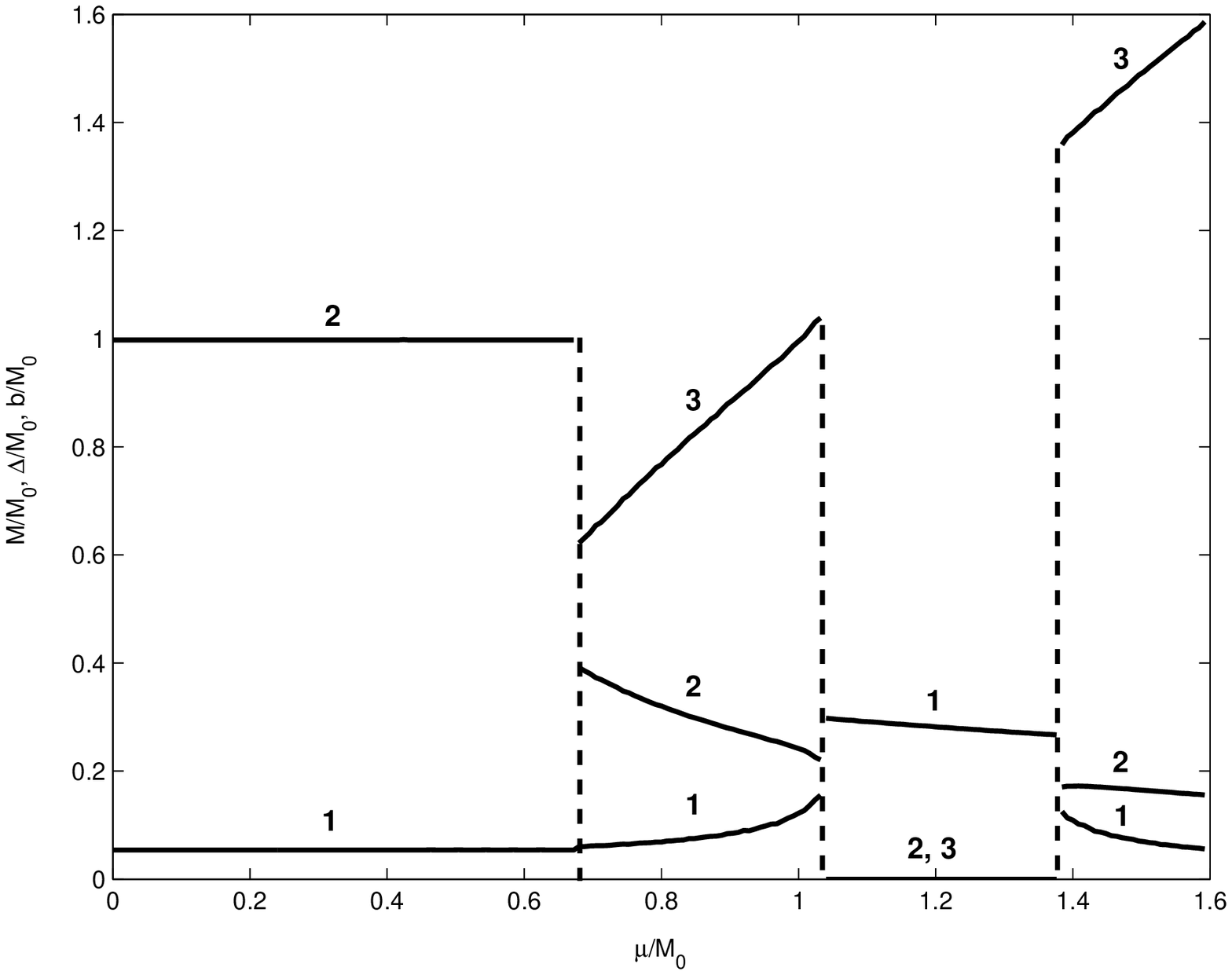}\\
\parbox[t]{0.45\textwidth}{
\caption{Phase portrait of the model when inhomogeneity of pion
  condensates is taken into account. Here IPC$_{1,2}$ denote
  inhomogeneous charged pion condensation phases, $n_q$ is a quark
  number density, 2 is a normal quark matter phase with $\Delta=0$,
  $M\ne 0$ (see in Fig. 1).} }
\hfill
\parbox[t]{0.45\textwidth}{
\caption{Gaps $M$ (line 1), $\Delta$ (line 2) and inhomogeneity wave vector $b$ (line 3) vs $\mu$ at fixed
$\nu=1.2M_0$.} }
\label{fig:Condensates_massive}
\end{figure}

We have studied numerically the TDP (\ref{21}) as a function of $M$,
$\Delta$ and $b$ for some physically motivated value of the massive parameter $m=M_0\tilde\alpha_0/\pi$, where $\tilde\alpha_0\approx 0.17$ and $M_0$ is a free parameter of the model (see the corresponding explanation at the end of the previous section III). The properties of its global minimum point vs
chemical potentials give us the phase structure of the model which
is presented in Fig. 3. It is easy to see that the phase structure in
the case of spatially inhomogeneous condensates (see Fig. 3) is the
same as in the case of homogeneous ones (see Fig. 1) but with two
essential exceptions. Namely, the normal quark matter phases 1 and 3
of Fig. 1 are replaced by two inhomogeneous pion condensation phases
IPC$_{1,2}$ in Fig. 3. The behavior of the gaps $M$ and $\Delta$ as
well as of the wave vector $b$ vs chemical potentials in these phases
are presented in Figs 4-6.
\begin{figure}
 \includegraphics[width=0.45\textwidth]{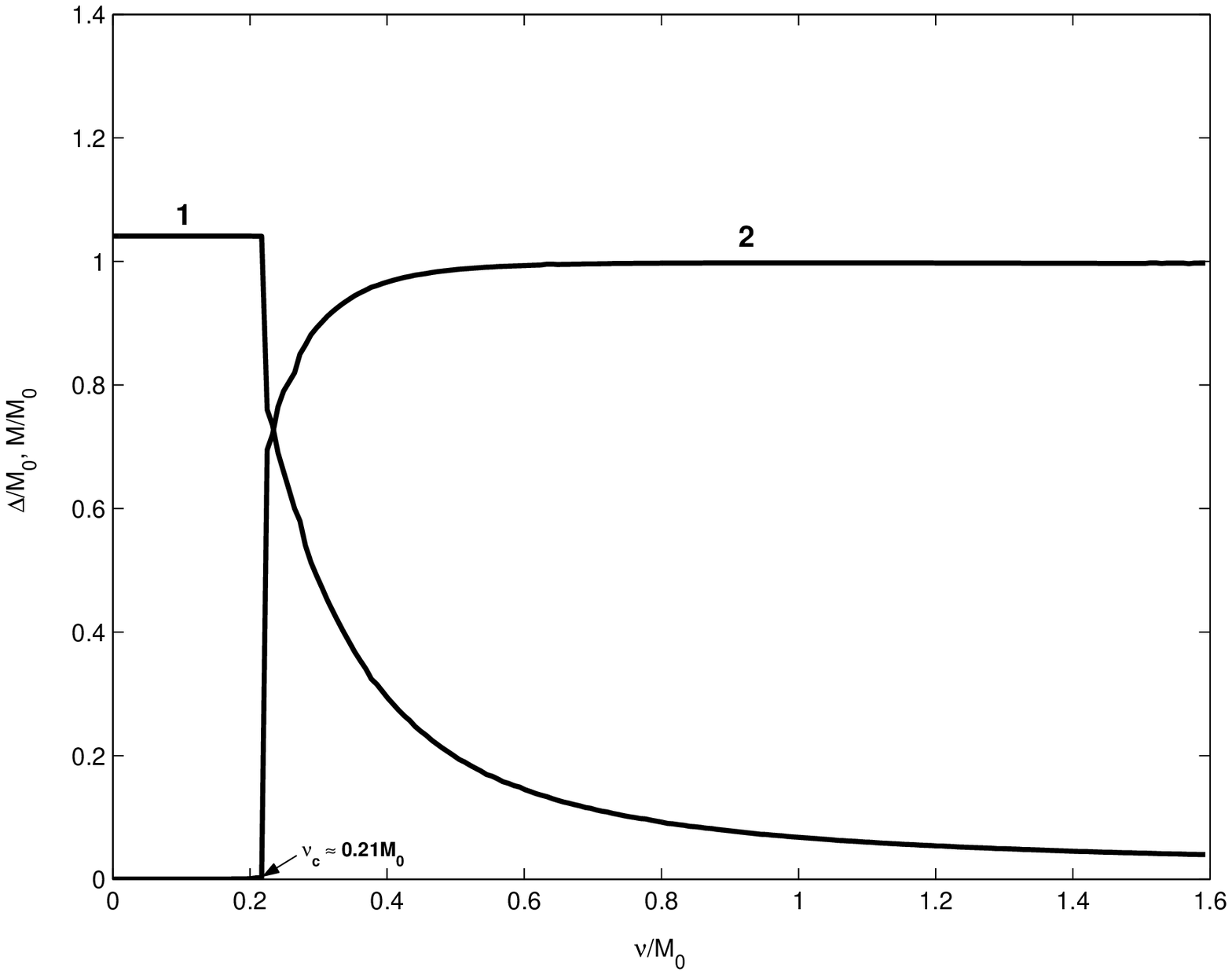}
 \hfill
\includegraphics[width=0.45\textwidth]{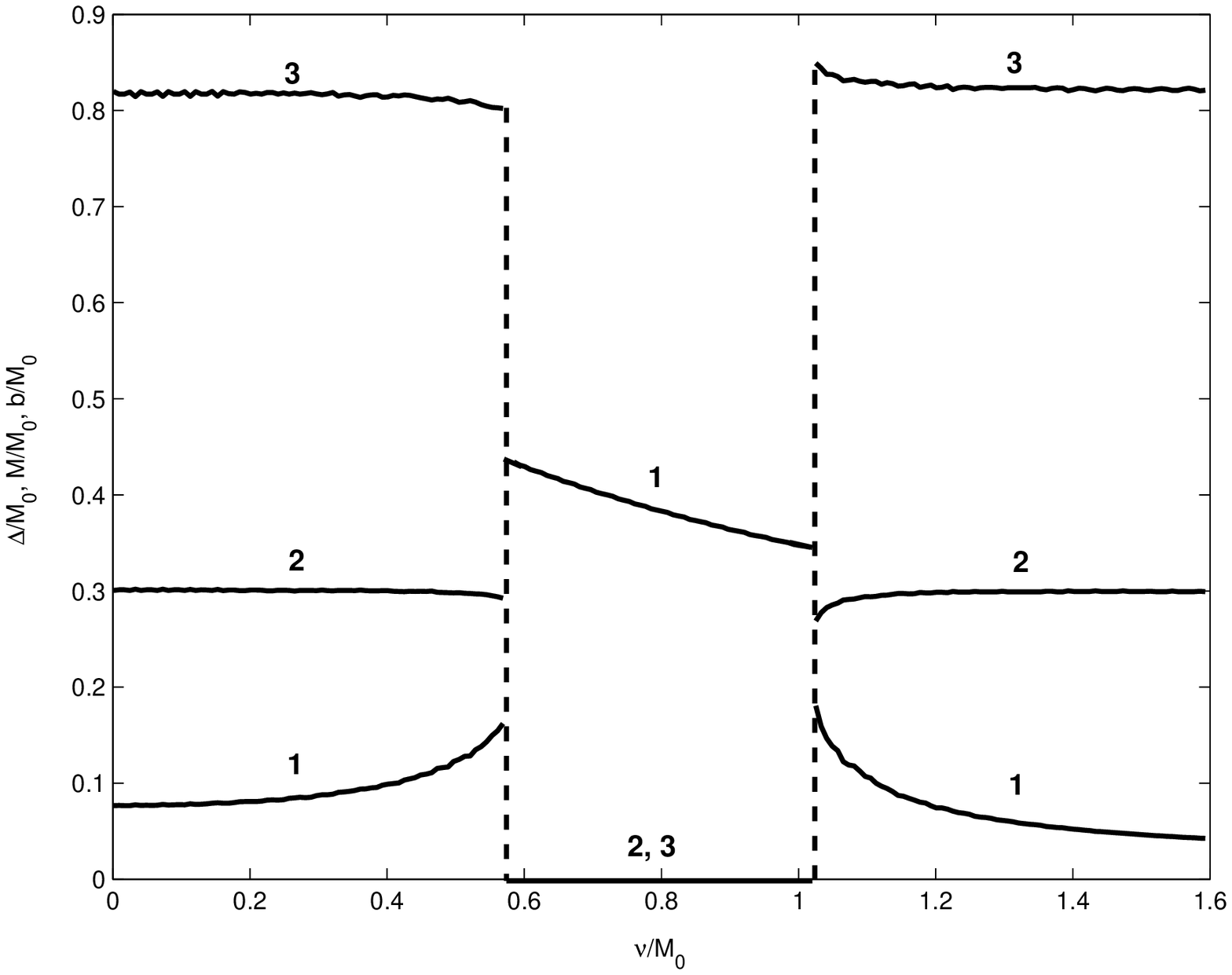}\\
\parbox[t]{0.45\textwidth}{
\caption{Gaps $M$ (line 1) and $\Delta$ (line 2) vs $\nu$ at fixed
$\mu=0.3M_0$. In this case $b\equiv 0$.} }
\hfill
\parbox[t]{0.45\textwidth}{
\caption{Gaps $M$ (line 1), $\Delta$ (line 2) and inhomogeneity
  wave vector $b$ (line 3) vs $\nu$ at fixed
$\mu=0.85M_0$.  }
}
\end{figure}
The TDP (\ref{21}) provides also the expressions for the quark
number density $n_q$ and isospin density $n_I$,
\begin{eqnarray}
n_q=-\frac{\partial\Omega^{ren}(M,b,\Delta)}{\partial\mu},~~~~
n_I=-\frac{\partial\Omega^{ren}(M,b,\Delta)}{\partial\mu_I}.
\label{33}
\end{eqnarray}
In particular, as it follows from (\ref{33}) and (\ref{21})
\begin{eqnarray}
n_q&=&\int_{-\infty}^\infty\frac{dp_1}{2\pi}\left [\theta
(\mu-\eta^{+}_1)+ \theta (\mu-\eta^{-}_1)+\theta
(\mu-\eta^{+}_2)+\theta (\mu-\eta^{-}_2)-2\right ]\nonumber\\&=&\int_{-\infty}^\infty\frac{dp_1}{2\pi}\left [\theta
(\mu-\eta^{+}_1)+ \theta (\mu-\eta^{-}_1)\right ],
\label{34}
\end{eqnarray}
where the last equality appeares due to the fact that $\eta^{\pm}_2<0$.
Using these expressions one can easily prove that in the
inhomogeneous pion condensation phases IPC$_{1,2}$ both isospin
density $n_I$ and quark number density $n_q$ are nonzero (see
Fig. 3). In contrast, in the homogeneous pion condensation phase of
Fig. 3 the density $n_q$ is zero. The behavior of $n_I$ and $n_q$ at some
particular values of chemical potentials are presented in Figs 7,
8. Hence, we have proved that in the
framework of the initial model the charged PC phenomenon with
nonzero $n_q$-density is possible only with spatially inhomogeneous
charged pion condensate.
\begin{figure}
 \includegraphics[width=0.45\textwidth]{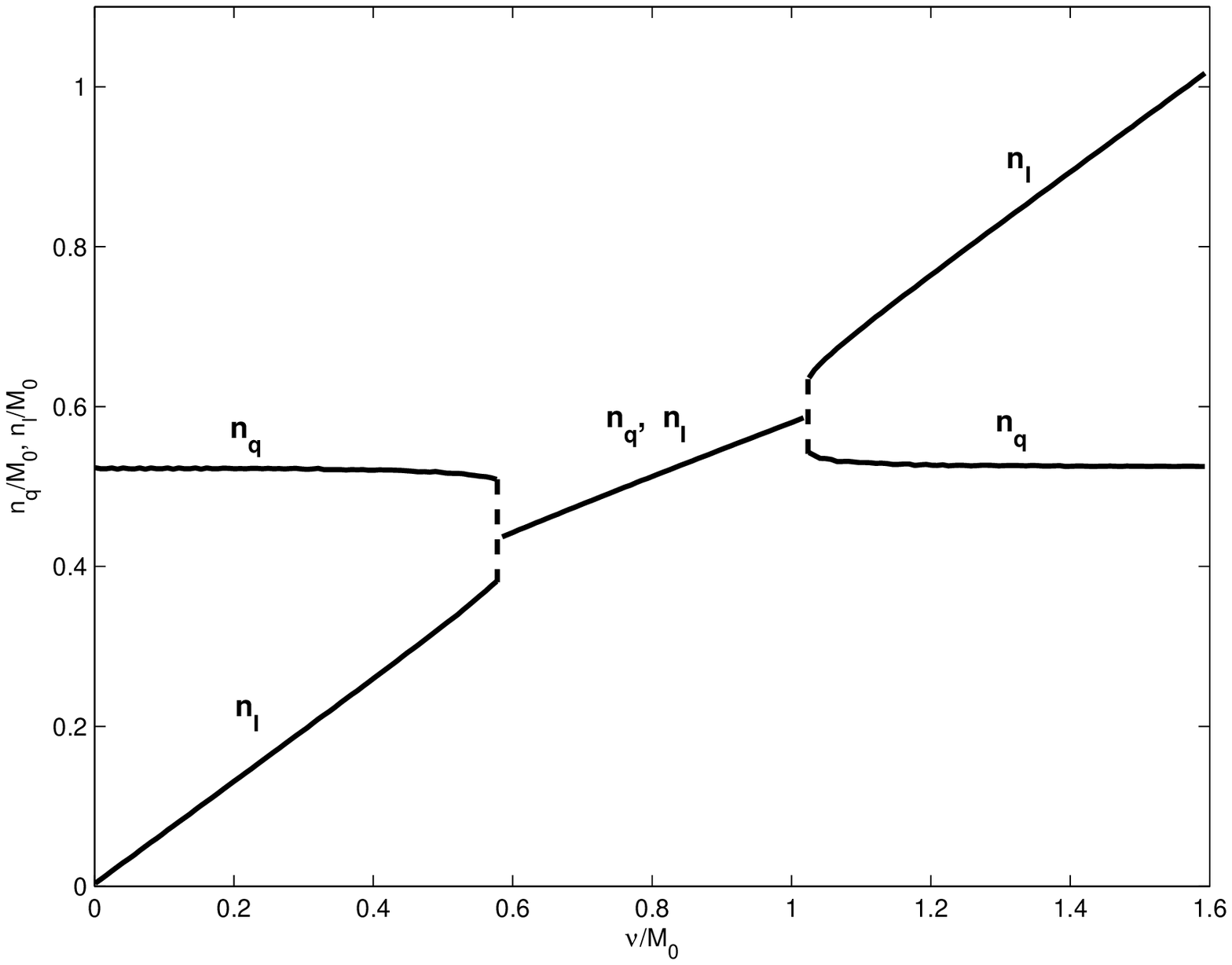}
 \hfill
\includegraphics[width=0.45\textwidth]{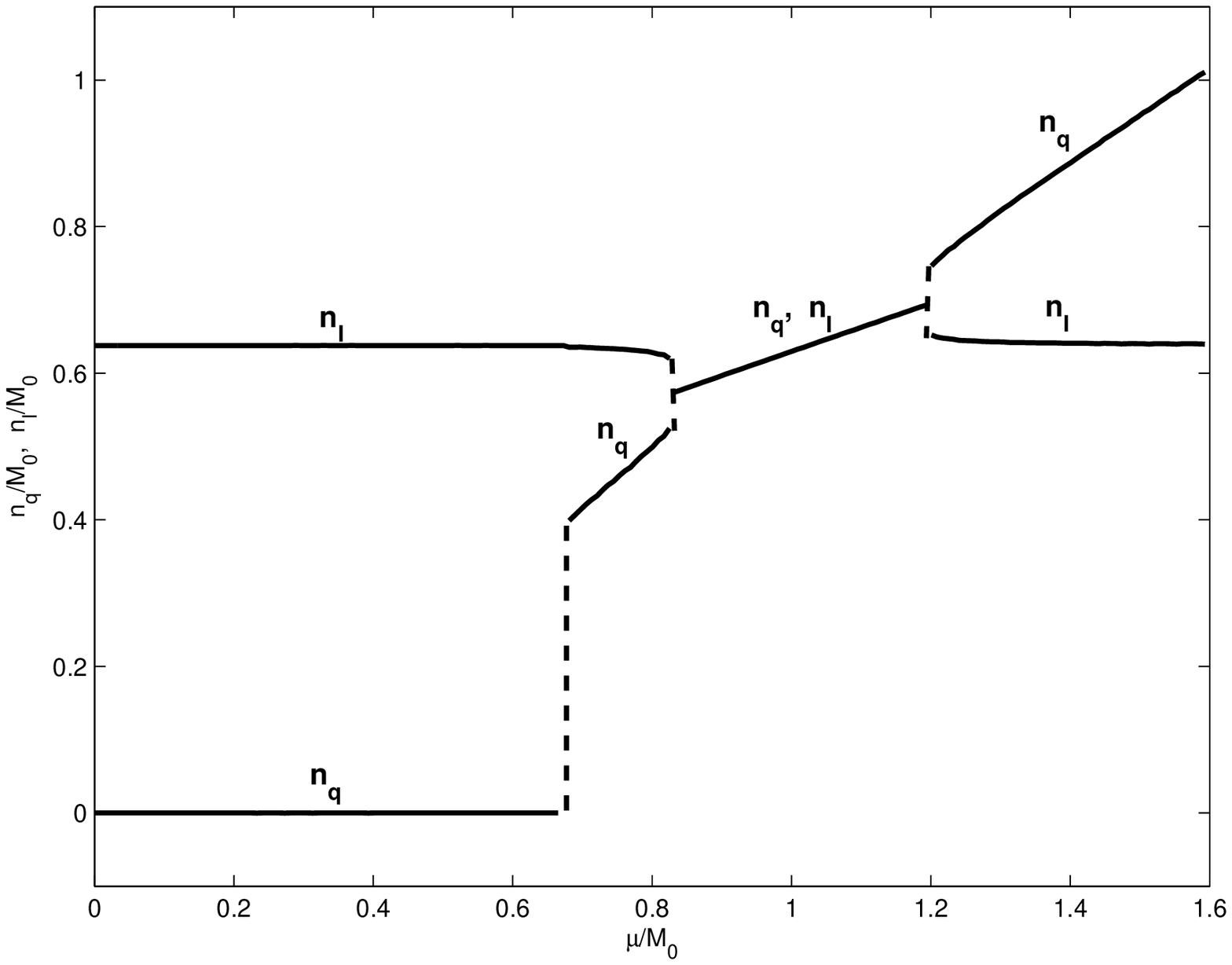}\\
\parbox[t]{0.45\textwidth}{
\caption{Quark number, $n_q$, and isospin, $n_I$, densities vs $\nu$ at $\mu=0.85M_0$.} }
\hfill
\parbox[t]{0.45\textwidth}{
\caption{Quark number, $n_q$, and isospin, $n_I$, densities vs $\mu$ at $\nu=M_0$.  }}\end{figure}

\section{Summary and conclusions}

This paper is devoted to the investigation of the so-called charged
pion condensation (PC) phenomenon which might be observed in dense
baryonic matter with different contents of $u$ and $d$ quarks. To
simplify the consideration, we have restricted ourselves to the
(1+1)-dimensional NJL-type model (1) with quark number $\mu$ and
isospin $\mu_I$ chemical potentials at zero temperature. Special
attention is paid to the influence of spatial inhomogeneity of
different condensates on charged PC phenomenon. Our consideration is
performed in the leading order of the large-$N_c$ expansion.

Recall, the charged PC phenomenon was studied recently in the
framework of some QCD-like effective theories such as NJL models or
chiral effective theories in the usual (3+1)-dimensional Minkowski
spacetime \cite{son,ek,ak,mu,andersen}. However, the existence of
the charged PC phase with {\it nonzero baryon or quark number
density}, denoted below as PCd phase, was there predicted  without
sufficient certainty. Indeed, for some values of model parameters
(the coupling constant $G$, cutoff parameter $\Lambda$ etc.) the PCd phase is allowed by NJL models. However, for other physically interesting
values of $G$ and $\Lambda$ the PCd phase is forbidden in the
framework of NJL models \cite{ek}. Moreover, if the electric charge
neutrality constraint is imposed,  the charged pion condensation
phenomenon  depends strongly on the bare (current) quark mass
values. In particular, it turns out that the PCd phase is forbidden
in the framework of NJL models if bare quark masses reach the
physically acceptable values of $5\div 10$ MeV (see in
\cite{andersen}).

As for  investigations of the charged pion condensation phenomenon in
the framework
of the (1+1)-dimensional massive/massless NJL model (1), the results  of
our recent papers show that the PCd phase is also absent there, if
PC condensate is  spatially homogeneous \cite{massive,gubina}
(see also Sec. III of the present paper). However, earlier we have found one factor which promotes the creation of the PCd phase at least in
(1+1)-dimensional spacetime. It is the finiteness of the volume of a
physical system \cite{tamaz}. Since such a constraint 
with certain boundary conditions imposed for any system is
equivalent to its consideration in a space with nontrivial topology,
we have studied in \cite{tamaz} the initial model (1) in the
spacetime $R^1\times S^1$ (spatial coordinate is
compactified) and proved that at some boundary conditions for spinor
fields the charged PCd phase is realized in the system.

In this paper, we have proved that a spatial inhomogeneity of
PC condensate is also a factor which promotes the appearance of PCd
phases on the phase diagram of the NJL$_2$ model (1). Indeed, if 
consideration of dense quark matter is performed in terms of
homogeneous pion condensates, then PCd phase is absent on the phase
diagram of the model (1) (see Fig. 1). However, if the spatial
modulation of pion condensates is taken into account in the form
(10), then  two PCd phases
appear on the phase diagram of the model (1) (those are IPC$_1$ and
IPC$_2$ phases of Fig. 3).

In summary, we conclude that charged pion condensation
phenomenon  of dense and isotopically asymmetric quark/hadron matter
is more preferable to be spatially inhomogeneous than homogeneous.

Finally, we would like to discuss the reliability of the  main result
of our paper and try to predict what might happen with the PCd phase in
the framework of a more general ansatz for condensates. To simplify
the problem, let us consider the case of massless NJL$_2$ model (1)
with  $m_0=0$, with an evident generalization of the ansatz (\ref{6}). Indeed, if $m_0=0$, then it is possible to study the phase
structure of the model in terms of the following simultaneous
spatial modulations of the chiral and pion condensates 
\begin{eqnarray}
\vev{\sigma(x)}=M\cos(2ax),~~~\vev{\pi_3(x)}=M\sin(2ax),~~~\vev{\pi_1(x)}=\Delta\cos(2bx),~~~ \vev{\pi_2(x)}=\Delta\sin(2bx). \label{60}
\end{eqnarray}
Evidently, there are three particular cases of (\ref{60}). i) The choice
$a=b=0$ correspongs to spatially homogeneous condensates and the phase
structure of the model for this parameterization was studied in
\cite{massive}. ii) Then, the choice $a=0$ is really the ansatz
(\ref{6}) at $m_0=0$. iii) Finally, the phase structure of the model
under the constraint $b=0$ was studied in \cite{gubina}. Recall that
in cases i) and iii) the PCd phase does not appear at the phase diagram. We have made preliminary estimations of the phase structure of the
massless model (1) in the framework of the ansatz (\ref{60}) and found
that 1) the phase with $M\ne 0$, $a\ne 0$, $\Delta\ne 0$, and $b\ne 0$
is absent. 2) There is an absolute minimum of the TDP corresponding to
spatially inhomogeneous PCd phase with $\Delta\ne 0$, $b\ne 0$, $M=
0$, $a= 0$. 3) For the same values of chemical potentials there is an
equivalent TDP extremum, corresponding to a chiral spiral phase, where
$M\ne 0$, $a\ne 0$, $\Delta= 0$, and $b= 0$. It means that inside an
inhomogeneous PCd phase, bubbles of the inhomogeneous phase with chiral spiral are allowed to exist and vice versa. By analogy, one might
expect that in the more physically interesting case
with $m_0\ne 0$, the spatially inhomogeneous charged PCd phase 
would continue to be present at the phase diagram of the model (1), even if
an arbitrary more general parameterization of condensates is used.

\appendix

\section{Roots of the equation $\eta^4 + A\eta^2 + B\eta + C=0$}
\label{ApA}

Using any program of analytical calculations, four roots of this
equation can  be presented in the following form:
\begin{eqnarray}
\eta_1^\pm=\frac 12\sqrt{P}\pm\left (-2A-P-\frac{2B}{\sqrt{P}}\right
)^{1/2},~~~~ \eta_2^\pm=-\frac 12\sqrt{P}\pm\left
(-2A-P+\frac{2B}{\sqrt{P}}\right )^{1/2}, \label{AA1}
\end{eqnarray}
where
\begin{eqnarray}
 P &=& -\frac{2A}{3} +
\frac{\sqrt[3]{2}~R}{3Q}
+\frac{Q}{3~\sqrt[3]{2}},~~~Q=\left(S+\sqrt{-4R^3+S^2}\right)^{\frac{1}{3}},\nonumber\\
R&=&A^2+12C,~~~~~~~S=2A^3+27B^2-72AC.
\end{eqnarray}
\section{Derivation of  formula (\ref{int})}
\label{ApB}

Let us denote the integral in the left hand side of (\ref{int})  by
$I$ (recall, there $a$ is a real quantity).

It well-known that in quantum field theory any loop
$p_0$-integration  is performed in the supposition that $p_0$ is a
shorthand notation for
$p_0+\mathrm{i}\varepsilon\cdot\mathrm{sign}(p_0)$, where
$\varepsilon\to 0_+$. In this case the causality of the theory is
preserved.  Taking this circumstance in mind, we see that in
(\ref{int}) the integration contour at $a>0$ ($a<0$) lies above
(below) the singularity point
$a-\mathrm{i}\varepsilon\cdot\mathrm{sign}(a)$ of an integrand
function. Hence, it is possible to perform in (\ref{int}) the Wick
rotation of the integration contour and to direct it along the
imaginary axis of the complex $p_0$-plane. In thus obtained integral
one can change an integration variable, $p_0\to\mathrm{i} p_0$. As a
result, we come to the relation
\begin{eqnarray}
I=\mathrm{i}\int_{-\infty}^\infty dp_0\ln\big (\mathrm{i}p_0-a)
=\mathrm{i}\int_{0}^\infty dp_0\ln\big (\mathrm{i}p_0-a)+
\mathrm{i}\int_{-\infty}^0 dp_0\ln\big (\mathrm{i}p_0-a). \label{B1}
\end{eqnarray}
In the last integral of (\ref{B1}) one can again change an
integration  variable, $p_0\to - p_0$. Hence,
\begin{eqnarray}
I=\mathrm{i}\int_{0}^\infty dp_0\ln\big (\mathrm{i}p_0-a)+
\mathrm{i}\int_0^{\infty} dp_0\ln\big (-\mathrm{i}p_0-a)=
\mathrm{i}\int_{0}^\infty dp_0\ln\big (p_0^2+a^2). \label{B2}
\end{eqnarray}
The last integral in (\ref{B2}) can be easily taken using  the
 integration by part method. Thus, up to an omitted infinite term which
does not depend on $a$, we  obtain $I=\mathrm{i}\pi |a|$.

\section*{Acknowledgments}

The authors  are grateful to Professor
D. Ebert for the support of our investigations and for many fruitful
discussions and to T.G. Khunjua for his interest to our problem.

\end{document}